\documentclass[preprint,aps,prd,amsmath,showpacs,nofootinbib]{revtex4}
\usepackage{graphicx}

\newcommand{\beq}{\begin{equation}}
\newcommand{\eeq}{\end{equation}}
\newcommand{\bea}{\begin{eqnarray}}
\newcommand{\eea}{\end{eqnarray}}

\newcommand{\Ref}[1]{(\ref{#1})}

\begin{document}

\title{"Gauge" freedom and relationship between the Einstein and Jordan conformal frames }

\author{Sergey M. Kozyrev}
\email{Sergey@tnpko.ru} \affiliation{Scientific center for gravity
wave studies ``Dulkyn'', Kazan, Russia}

\author{Rinat A. Daishev }
\email{Rinat.Daishev@ksu.ru} \affiliation{$^1$ Department of
Mathematics and Department of Physics, Kazan Federal University,
Kremlevskaya str. 18, Kazan 420008, Russia}

\begin{abstract}
The issue of the physical equivalence between the Einstein and
Jordan conformal frames in Jordan-Brans-Dicke (JBD) theory is
revised. Scalar-tensor theories equations are not invariant with
respect to conformal transformations if one uses the same "gauge"
fixing in Jordan and Einstein frames. We hope to have clarified
some eventual obscure issues associated to the concept of
conformal invariance of relativistic theories appearing in the
literature, in particular the relationship between Jordan and
Einstein frames.
\end{abstract}

\pacs{04.20.-q, 04.20.Jb, 04.50.Kd}

\maketitle

\section{Introduction}
One of the largest concentrations of literature within the area of
relativistic gravity theories is interpretations of exact
solutions of field equations. Some of them are discovered at the
early stage of development of relativistic theories, but up to now
they are often considered as equivalent representations of some
"unique" solution. On the other hand, the variables $x$, used in
Einstein's equations, represent co-ordinates of points of abstract
four-dimensional manifold $M^4$ over which there are a set pseudo
Rimanien spaces $V^4(g)$, generated by set of solutions
$g_{\alpha\beta}$. Co-ordinates in each of such spaces have the
specific properties differing from their properties in other
spaces \cite{Gullstrand}. Note that an affine connection which to
attach to gravity theory can be at most an independent postulat of
theory \cite{Brans}. In this case the point dependent property of
manifolds is linked with the fact that the units for measure of
underlying geometry are running units. For example, it could be a
theory based on Brans Dicke action but endowed with space time of
Weyl integrable structure \cite{Matos}.

 Evidently that needs preliminary a
clear mathematical distinction between the concepts of
co-ordinate, "gauge", reference frames in the relativistic
theories. To obtain the unknown components of metric tensor from
the field equations we must to do the following conventions:

First, sets of numbers, with whom we conduct manifold
arithmetization (e.g. spherically symmetric). Before solving field
equations we have only a differential manifold structure endowed
with an affine connection. The field equations of relativistic
theory can be derived from the action up to boundary terms. The
angles and distance are induced rather than fundamental concepts
in this proposal. It is important to note that the two geometrical
structures, the metric and arithmetization, are fundamentally
independent geometrical objects.

Second, from geometrical point of view one has to introduce an
additional mathematical structure - describing some specific
principle of construction of space-time model is responsible for
measuring the distances - the "gauge". On the other words "gauge"
is a rule for reception of "coordinate system" on a single
manifold $M$ (e.g. harmonic, isotropic, curvature coordinates).It
is important to realize that the field equations alone are not
enough to determine a gravitational system, while these equations
are a set of 6 nonlinear partial differential equations for the 10
metric components. Einsteins equations determine the solution of a
given physical problem up to four arbitrary functions, i.e., up to
a choice of "gauge". Evidently, a structure of space-times is
mathematically represented by Einsteins equations and four
co-ordinate conditions \cite{Temchin}, which considered
independent of the action
\begin{subequations} \label{BDeqw}
\begin{eqnarray}\label{BDeq}
&G_{\mu\nu}=T_{\mu\nu} ,& \\
&C(x)g_{\mu\nu} = 0,\label{BDeqI}&
\end{eqnarray}
\end{subequations}
 where $g_{\mu\nu}$ metric
tensor and $C(x)$ - some algebraic or differential operators.
Thereby for any four of components $g_{\mu\nu}$ emerge the
relations with remaining six and, probably, any others, known
functions. Certainly, equations \Ref{BDeqI} cannot be covariant
for the arbitrary transformations of independent variables, and
similarly should not contradict Einstein's equations or to be
their consequence. Moreover, four of ten field equations will not
be transformed according to any rules, but simply replaced by hand
with the new. This "gauge" is the unphysical degree of freedom and
we must fix the "gauge" or extract some invariant quantities to
obtain physical results \cite{Nakamura}.

Third, the unknown components of metric tensor $g_{\mu\nu}$  are
determined from the solutions of Einstein's field equations. (e.g.
Schwarzschild, Heckmann, Brans solutions). Consequently, the
geometrically interpreted co-ordinate system of obtained
space-time and any relationship it derives from equations
\Ref{BDeq}, \Ref{BDeqI} emerge a posteriori \cite{Temchin}.
Moreover, property of this co-ordinate system will depend from
initial and boundary conditions for \Ref{BDeq}, \Ref{BDeqI}. An
intriguing consequences of the above discussion is the "gauge"
freedom can be expected in relation with some connection to
problems in quantum physics. Generally speaking, occurrence of the
observer ("gauge" fixing) influences results of measurements and
physics are different in two different "gauges".

\section{Conformal transformation (Weyl rescaling)}

    The important feature of the JBD gravity is connected with the
conformal symmetry. It is well known, since the pioneering paper
of Jordan \cite{Jordan} that the action of a scalar tensor theory
is invariant under local transformations of units that are under
general conformal transformations, or sometimes called Weyl
rescaling:

\begin{equation}
 g_{\mu\nu}\rightarrow \widehat{g}_{\mu\nu}=\Omega^{2}(x) g_{\mu\nu},
 \quad \mbox{or}\quad ds^2 \rightarrow d\widehat{s}^2 = \Omega^{2}(x)ds^2.
\label{BDeq2}
\end{equation}
where $\Omega(x)$ a local arbitrary function of $x$.

 This method of conformal transformation provides a clear and powerful technique, free
from mathematical ambiguity, but nevertheless requires careful
consideration from the physical point of view.

Among all conformally related frames one distinguishes two frames:
Jordan's and Einstein's. Let us consider the pure gravitational
sector of the JBD theory (as a minimal extension of general
relativity) in the Jordan's conformal frame, the field equations
\Ref{BDeq} can be derived from the following action

\begin{equation}
L(g,\phi)=\sqrt{-g}\left(\phi R - \frac{\omega}{\phi}g^{\mu\nu}
\nabla_\mu \phi \nabla_\nu \phi \right) +
L_{matter}(g)\label{BDeqs}
\end{equation}
where $R$ is the curvature scalar, $\phi$ is the scalar JBD field,
$\omega$ is the JBD coupling constant, and $L_{matter}[g]$ is the
Lagrangian density of the ordinary matter minimally coupled to the
scalar JBD field.

 The graviational part of the Jordan's frame JBD Lagrangian
density $L(g,\phi)$ \Ref{BDeqs} without ordinary matter is
invariant in form under the conformal rescaling of the spacetime
metric \cite{Jordan}, \cite{far}:

\begin{equation}
g_{\mu\nu} \rightarrow
\widehat{g}_{\mu\nu}=\phi^{2\alpha}g_{\mu\nu}\label{BDcon}
\end{equation}

 After the conformal transformations \Ref{BDcon},
we get Lagrangian for the Einstein's frames

\begin{equation}
L(\hat g,\hat \phi)=\sqrt{-\hat g}\left(\hat R - (\omega +
\frac{3}{2}) \widehat{g}^{\mu\nu} \widehat{\nabla}_\mu
\widehat{\phi} \widehat{\nabla}_\nu \widehat{\phi}\right) +
\widehat{L}_{matter}( \widehat{g}, \widehat{\phi} ) \label{eq3}
\end{equation}
The scalar function $\hat \phi \equiv \ln \phi$ is the JBD scalar
field in the Einstein frame, and $\hat L_{matter}(\hat g,\hat
\phi)$ is the Lagrangian density for the ordinary matter
nonminimally coupled to the scalar field.

Note the possibility of changing the coupling in \Ref{BDeqs};
$L_{matter}(g) \rightarrow L_{matter}(g,\phi)$, while keeping
intact the gravitational part:

\begin{equation}
L^{*}(g,\phi)=\sqrt{-g}\left(\phi R - \frac{\omega}{\phi}
g^{\mu\nu} \nabla_\mu \phi \nabla_\nu \phi\right) +
L_{matter}(g,\phi)
\end{equation}

In this case the ordinary matter is nonminimally coupled to the
scalar field in the Jordan frame.

In addition one can obtain the Lagrangian density to the Einstein
frame in the form:

\begin{equation}
L^{*}(\hat g,\hat \phi)=\sqrt{-\hat g}\left(\hat R - (\omega +
\frac{3}{2}) \widehat{g}^{\mu\nu} \widehat{\nabla}_\mu
\widehat{\phi} \widehat{\nabla}_\nu \hat \phi\right) + \hat
L_{matter}(\hat g)
\end{equation}
In this case the scalar field $\hat \phi$ is minimally coupled to
the curvature. Note that, unless a clear statement of what is
understood by "equivalence of frames"- is made, the issue which is
the physical conformal frame is a semantic one. Hence, there are
four related but inequivalent scalar-tensor theories in Jordan and
Einstein frame \cite{Quiros}.

 In the literature, the physicists do not agree with
each other about the equivalence of the two frames (see review in
\cite{FM}). However, the meaning of the equivalence between the
Jordan frame and the Einstein frame is not assuming the additional
equations \Ref{BDeqI}. These equations put by hand and not
covariant.  This issue is critical for the interpretation of the
predictions of a given theory of gravity since these seem to be
deeply affected by the choice of the coordinate conditions
\cite{Gullstrand}. For concreteness, let us consider "harmonic
gauge" of coordinate \cite{Fock},

\begin{eqnarray}\label{CHarm}
g_{\mu\nu} \Gamma^{\alpha}_{,\mu\nu}=0.
\end{eqnarray}
which usually assumed as the analogue of Lorenz gauge, $\partial
A=0$, in electromagnetism. However this analogy is the most
superficial: this or that gauge in nonrelativistic theory is a
problem of exclusively convenience, it's this or that expedient
does not influence in any way on a values of physical quantities
and it is not related to observation requirements, - whereas the
choice of co-ordinate system is related to all it essentially.

In fact, there are the related but inequivalent scalar-tensor
theories in Jordan and Einstein frame. The reason is very simple.
If we use the same conformal transformations, like the
\Ref{BDcon}, in both the equations \Ref{BDeq} and \Ref{BDeqI},
then the in and out states are not the same in the two frames.  If
one postulates that the field equations are invariant with respect
to conformal transformations \Ref{BDeqI}, one obtains in addition
transformations of co-ordinate conditions \Ref{CHarm}

\begin{eqnarray}\label{CHarm}
g_{\mu\nu} \Gamma^{\alpha}_{,\mu\nu}=\widehat{g}_{\mu\nu}
\widehat{\Gamma}^{\alpha}_{,\mu\nu}+\frac{\partial_\mu
\Omega}{\Omega}.
\end{eqnarray}
As a result, since the Einstein field equations are undetermined;
scalar-tensor theories cannot achieve the harmonic metric for any
$\Omega$ functions but only when $\Omega$  is taken a constant.
One must assume that two frames represent not the same set of
physical gravitational and non-gravitational fields. In fact, two
conformally connecting spaces $V^4(g)$ and $\widehat{V}^4(g)$ are
given not in the same manifold. Consequently, under this conformal
transformation the solution of some initial physical problem will
be transformed onto a solution of a completely different problem.
Thus, applying the same coordinate conditions in different
physical requirements, we arrive at dissimilar physical theories,
because we are solving different equations.

On the other hand each scalar-tensor theory can be transformed
into general relativity plus conformally invariant scalar field
\cite{Sokolowski}. The gravitational interaction for scalar tensor
theories is taken into account by the Einstein equations, which
are generally written in the form  \Ref{BDeq}. The Einstein tensor
$G_{\mu\nu}$ is constructed from the geometrical properties of the
space-time, while $T_{\mu\nu}$  is the energy momentum tensor of
matter. One can in principle assume gauge-dependence of
right-hand-side of equation \Ref{BDeq} as a variety of matter
fields with different equations of state. Now, if we consider, the
system \Ref{BDeq}, \Ref{BDeqI} as equations for same "gauge"
fixing then the Jordan's and Einstein's conformal frames can be
viewed as a different "matter source" of energy momentum tensors
$T_{\mu\nu}$ of Einstein's equations. The physical content of this
point of view can be stated in the following simple way: the
equations of state for "matter" are not the same for the different
conformal frames is chosen.

It is evident that in different conformal frame representations of
JBD theory are neither mathematically, nor physically equivalent.

\section{Conclusion}

In this article, we clarify the notion of "gauges" in relativistic
theories, which is necessary to understanding the physical
equivalence between the Einstein and Jordan conformal frames. We
have shown the procedure to find gauge-invariant variables in the
scalar tensor theories through the precise treatments of "gauges".
In method of conformal transformation, we always treat two
space-time manifolds. One is the space-time for Jordan frame and
the other is the space-time for Einstein frame. Note that these
two space-times for Jordan and  Einstein frame are distinct. In
any relativistic theories, we must always write additional four
co-ordinate conditions \Ref{BDeqI}. These "gauges" may describe
different physical solutions of Einstein equations with the same
space arithmetization. The conformal transformations are not
diffeomorphisms of the single manifold $M$, and the transformed
metric $\widehat{g}_{\mu\nu}$ is not simply the metric
$g_{\mu\nu}$ written in a different coordinate system these
metrics describe different gravitational fields and different
physics.

Keeping in our mind that we always treat two different space-times
in Jordan and  Einstein frame. Eq. \Ref{BDeqI} is a rather curious
equation because it not covariant for the arbitrary
transformations of independent variables. In this case the metric
is left unchanged, although its coordinate representation varies.
In short, Eq. \Ref{BDeqI} gives a relation between variables on
two different manifolds.

\end{document}